# Imaging Biomarkers for Neurodegenerative Diseases from Detailed Segmentation of Medial Temporal Lobe Subregions on *in vivo* Brain MRI Using Upsampling Strategy Guided by High-resolution *ex vivo* MRI

Yue Li[1], Pulkit Khandelwal[1], Long Xie[2], Laura E. M. Wisse[3], Amanda E. Denning[1], Christopher A. Brown[1], Emily McGrew[1], Sydney A. Lim[1], Niyousha Sadeghpour[1], Sadhana Ravikumar[1], Ranjit Ittyerah[1], Eunice Chung[1], Daniel T. Ohm[1], Nidhi S. Mundada[1], María Mercedes Íñiguez de Onzoño Martín[4], María del Mar Arroyo Jiménez[4], Monica Mũnoz[4], Maria del Pilar Marcos Rabal[4], David J. Irwin[1], Edward B. Lee[1], Ricardo Insausti[4], Sandhitsu R. Das[1], David A. Wolk[1], Paul A. Yushkevich[1]

[1]University of Pennsylvania, Philadelphia, USA

[2] Siemens Healthineers, Princeton, USA

[3] Lund University, Lund, Sweden

[4] University of Castilla La Mancha, Albacete, Spain

yue.li@pennmedicine.upenn.edu

**Abstract.** The medial temporal lobe (MTL) is a region impacted extensively and nonuniformly in early stages of Alzheimer's disease (AD). Regional MTL morphometric measures extracted from magnetic resonance imaging (MRI) are supportive features for the diagnosis of AD and related disorders (ADRD). Different MRI modalities have distinct advantages for MTL morphometry. Anisotropic T2-weighted (T2w) MRI is preferred for hippocampal subfields due to its higher contrast between hippocampal layers. Isotropic T1-weighted (T1w) MRI is beneficial for thickness calculation of extra-hippocampal subregions due to its stable image quality and isotropic resolution. We propose a multi-modality MTL segmentation algorithm that bridges the T1w and T2w modalities by bringing both to a nearly isotropic voxel space. Guided by high-resolution *ex vivo* 9.4T MRI, an upsampling model was designed for the ground truth segmentations. Combined with non-local means upsampling, this model was used to construct a nearly isotropic T1w and T2w MTL subregion segmentation training set, which was used to train a nnUNet model. Morphometric biomarkers extracted by this model were compared to those extracted using conventional models operating in anisotropic spaces on downstream tasks. Biomarkers extracted using the proposed model had greater ability to discriminate between individuals with mild cognitive impairment and cognitively unimpaired; and had greater longitudinal stability. These findings suggest that the biomarkers derived from T1w and T2w MRI unsampled to nearly isotropic resolution have significant potential for improving disease diagnosis and monitoring disease progression in ADRD.

## 1 Introduction

Structural magnetic resonance imaging (MRI) biomarkers are crucial supporting features for diagnosing neurodegenerative diseases such as Alzheimer's disease (AD) [1] and serve as important tools for monitoring disease progression [2]. The medial temporal lobe (MTL), consisting of the hippocampus, amygdala, and parahippocampal gyrus, is a region of special significance in AD because it is affected early and extensively; and because patterns of change in MTL subregions can help distinguish between AD and other neurodegenerative diseases that have similar impacts on cognition but distinct underlying pathologies [3, 4].

Accurate segmentation of MTL subregions is a necessary step for extracting MRI biomarkers such as thickness and volume. Well-established automated segmentation approaches include ASHS [5], FreeSurfer [6], and HippUnfold [7]. Several papers have argued that the ~1 $mm^3$ isotropic T1-weighted (T1w) scans acquired almost universally in AD MRI studies lack sufficient detail and contrast for segmenting subfields of the hippocampus (which include cornu Ammonis fields 1-3 (CA1-CA3), dentate gyrus (DG), subiculum (SUB)) [8, 9]. For manual segmentation of hippocampal subfields, most published papers utilize anisotropic T2-weighted (T2w) scans ($\sim 0.4 \times 0.4 \times 2$ $mm^3$, where the third dimension represents the anterior–posterior direction) aligned with the major axis of the hippocampus to optimize visualization of hippocampal layers. However, T1w offers its own advantages, with isotropic resolution allowing better extraction of complex folding geometry of the parahippocampal gyrus; and greater robustness to participant head motion [10]. Although both T1w and T2w scans can be utilized in ASHS, FreeSurfer, and HippUnfold, multi-modality data are primarily used in preprocessing steps, such as registration and region of interest determination, while the segmentation step focuses on a single modality. As a result, the advantages of each modality are not fully integrated.

Multi-modality segmentation has the potential to optimally combine the strengths of different imaging modalities. A multi-modality segmentation model, which can handle MRI scans with different field strengths and contrasts from the same participant, for MTL subregions was proposed in [11]. While this model effectively combines the advantages of contrast and quality of different MRI modalities, it aligns and resamples all available modalities into the space of a "primary" modality in its preprocessing step, resulting in image degradation. For example, when T2w is the primary modality, T1w images are resampled into T2w space, and the advantages of isotropic voxels are lost. Isotropic voxel spacing is important for AD biomarker extraction, especially for thickness calculation in the parahippocampal gyrus, whose accuracy relies on the voxel spacing.

While the problem of loss of T1w resolution can be trivially addressed by upsampling the T2w image (e.g., to $0.4 \times 0.4 \times 0.4$ $mm^3$) using super-resolution methods, a difficulty arises because the ground truth manual segmentations needed to train automatic methods are only available in native T2w resolution. To address this problem, we developed a cross-scale manual segmentation upsampling method guided by high-resolution *ex vivo* MRI. Using this method, we upsample an existing training set consisting of T2w MRI, T1w MRI and T2w-space manual segmentations to a common "nearly isotropic" space; and use this upsampled training set to train a multi-modality nnUNet model for MTL subregion segmentation. The MTL morphological measures extracted by this model enhance discrimination of patients with mild cognitive impairment (MCI) from cognitively unimpaired (CU) adults in cross-sectional analyses relative to existing approaches and demonstrate a greater temporal stability in longitudinal follow-up.

## 2 Methods

Figure 1 illustrates the flowchart of the proposed pipeline. In step 1, a cross-scale upsampling model is trained (Section 2.3). In step 2, using the trained upsampling model, a multi-modality atlas is upsampled to a high-resolution space, enabling the training of a nnUNet segmentation model (Section 2.4). In step 3, the segmentation method is applied to two independent test sets, and the accuracy and stability of the imaging biomarkers are verified (Section 2.5).

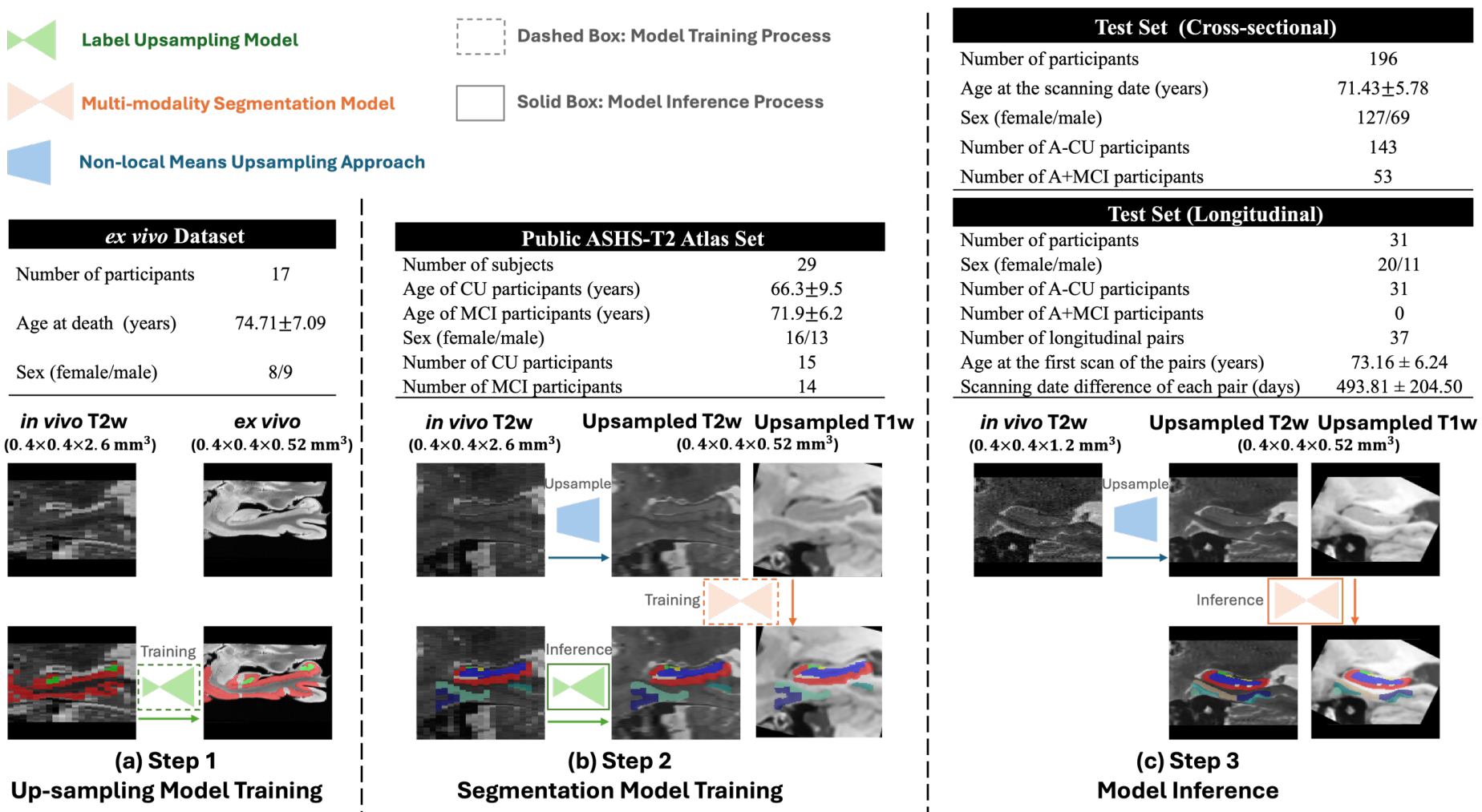

**Figure 1.** The flowchart of proposed nearly isotropic subregion segmentation model for medial temporal lobe. *Abbreviations*: CU, cognitively unimpaired; MCI: mild cognitive impairment; A+: positive amyloid-β status; A-: negative amyloid-β status.

### 2.1 Definition of Nearly Isotropic Space

The high-resolution space explored in this study is based on the resolution of the T2w images in the training set (most with a voxel spacing of $0.4 \times 0.4 \times 2.6$ mm$^3$). To unify the resolution in all directions, a discrete upsampling by an odd multiple of K is performed along the anterior–posterior direction and that K=5 is the value that made the upsampled images as close to isotropic as possible ($0.4 \times 0.4 \times 0.52$ mm$^3$). This upsampled space is referred as nearly isotropic space.

### 2.2 Dataset and Preprocessing

***ex vivo* training set for label upsampling model.** High-resolution ($0.2 \times 0.2 \times 0.2$mm$^3$) 9.4T *ex vivo* MRIs were collected from 17 participants (Figure 1(a)). MTL subregions were segmented manually based on histology images that had been co-registered to the *ex vivo* MRI and had been annotated by a team of neuroanatomists [12]. In addition, antemortem *in vivo* T2w and T1w MRIs of the same 17 individuals were available. To register the *ex vivo* MRI to the *in vivo* space, the T1-ASHS pipeline [10] was applied to the *in vivo* T1w MRI. Both sets of segmentations (detailed *ex vivo*, automated *in vivo* T1w) were reduced to just hippocampus label and extra-hippocampal MTL label, and rigid followed by deformable registration was performed to match each participant's *ex vivo* MRI to their *in vivo* MRI space. After registration, *ex vivo* MTL segmentations were transferred onto the *in vivo* T2w MRI (which had $0.4 \times 0.4 \times 2.6$ mm$^3$ resolution for most participants) followed by extensive manual editing to account for residual registration errors [13]. Ultimately, for these 17 participants the same set of anatomical labels was available in both *in vivo* T2w and *ex vivo* space. For the purposes of training the label upsampling network (Section 2.3), the *ex vivo* MRI was resampled into the "nearly isotropic" space ($0.4 \times 0.4 \times 0.52$ mm$^3$ for most cases); and various anatomical labels were combined into just two: a DG label (green) and MTL gray matter label (red), highlighted in Figure 1(a).

***in vivo* training set for multi-modality subregion segmentation model.** The public ASHS-T2w atlas [5] collected from 29 participants was utilized to develop the multi-modality segmentation model. The demographic information for these participants is displayed in Figure 1(b). All participants in the atlas had both T1w and T2w images, along with manual segmentations performed on the T2w images. The resolution of the T1w images is $1 \times 1 \times 1$ mm$^3$ while that of most T2w images is $0.4 \times 0.4 \times 2.6$ mm$^3$. The annotated labels included CA1-3, DG, SUB, entorhinal cortex (ERC), Brodmann areas 35 (BA35), Brodmann area 36 (BA36), and parahippocampal cortex (PHC).

**Independent test set.** Two test datasets were collected from a cohort that is independent of the two training sets mentioned earlier (Figure 1(c)). First, two groups with 196 participants, where differences in MTL morphometry due to AD were expected, were collected to form the cross-sectional test set. These two groups are: MCI patients who had tested positive for amyloid-β (A+), and CU participants who had tested negative for amyloid-β (A-). Another test set is a longitudinal dataset involving 31 A- CU individuals. Each participant underwent at least two MRI scanning sessions, with the scanning dates of adjacent sessions limited to within two years of each other. Throughout the follow-up period, the diagnoses and amyloid-β status of the participants remained unchanged. In total, 37 longitudinal pairs of scans were collected. The image resolutions of the two test sets differ from those of the training sets. The resolution of the T2w images within the coronal plane is $0.4 \times 0.4$ mm$^2$, while the coronal planes thickness ranges from 1.2, 1.4, 1.6, and 2 mm. The T1w images are isotropic, with resolutions of $0.8 \times 0.8 \times 0.8$ mm$^3$ and $1 \times 1 \times 1$ mm$^3$.

### 2.3 Label Upsampling Network

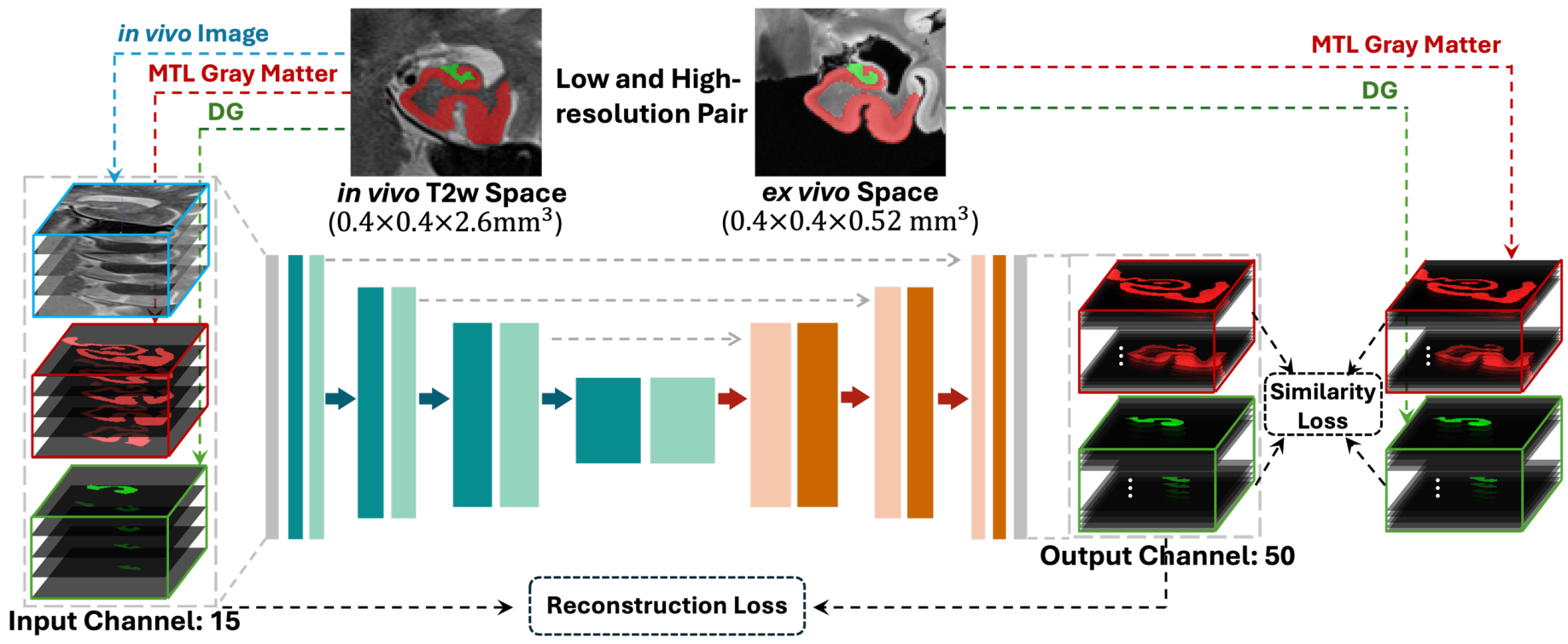

**Figure 2.** The architecture of the label upsampling network. *Abbreviations*: DG, dentate gyrus

In the *ex vivo* dataset, a label upsampling network is trained. This network is a 2D U-Net [14] (Figure 2) with 15 input channels and 50 output channels. The network consists of four scale levels, each with channel dimensions of 80, 160, 320, and 640, respectively. Within each scale level, there are two residual unit blocks. Downsampling between adjacent non-bottom levels is achieved using the first residual unit, which has a stride of 2.

During training, the network receives input in the form of coronal image patches sized 32×32 pixels. The center of each patch is randomly selected within the segmentation region from the *in vivo* image space. The same region is also cropped from the upper and lower two coronal slices. In total, each patch consists of 15 channels, including five original coronal slices from the *in vivo* images and their corresponding binary segmentations for the DG and MTL gray matter.

The network outputs a 50-channel result that represents the predicted one-hot label map for the segmentation, upsampled by a factor of five along the anterior–posterior axis. The first 25 channels indicate the gray matter label, while the remaining 25 channels represent the DG label.

The ground truth for the upsampled label is obtained from manual segmentation of high-resolution *ex vivo* images. The training process is guided by two loss functions. The first is the similarity loss ($L_{sim}$), a Dice loss that measures the similarity between the upsampling results and the ground truth. The second is the reconstruction loss ($L_{rec}$), another Dice loss that compares the average-pooled predicted segmentation (pooled with a factor of five reduction along the anterior–posterior axis) to the original *in vivo* segmentation. The total loss is the weighted sum of them: $L_{total} = 2 \cdot L_{sim} + L_{rec}$.

### 2.4 Multi-modality Segmentation Model

In the public *in vivo* ASHS-T2 atlas, the multi-modality high-resolution segmentation model is trained. To achieve this, we first need to upsample low-resolution multi-modality atlases. This involves two steps: upsampling MR images and upsampling manual segmentations. As shown in Figure 1(b), the non-local means method [15] is used for image upsampling. The T2w images are upsampled with a factor of five along the anterior–posterior axis to form a nearly isotropic space. The corresponding T1w images are upsampled to the same space using non-local means and linear interpolation.

For the manual segmentations, the data is upsampled through four steps. First, all gray matter labels except for DG in the T2w space are merged into a single MTL gray matter label (red label in Figure 3(b)). Second, gray matter label, DG label and the original T2w image are fed into the trained label upsampling network, which outputs the probability maps for upsampled gray matter (Figure 3(c)) and DG (Figure 3(d)), respectively. Third, the fast-marching algorithm computes the distances between voxels in the upsampled space and the different subregion labels from the original manual segmentation, allowing each voxel to be assigned to its closest label and creating a per-label posterior probability stack. A SoftMax operation and a voting scheme are used on the probability stack to generate the discrete label map (Figure 3(e)). Fourth, a cerebrospinal fluid (CSF) correction model, trained using a two-dimensional standard nnU-Net [16] in coronal view, is utilized to correct the CSF region that is covered by the surrounding gray matter labels (Figure 3(f)).

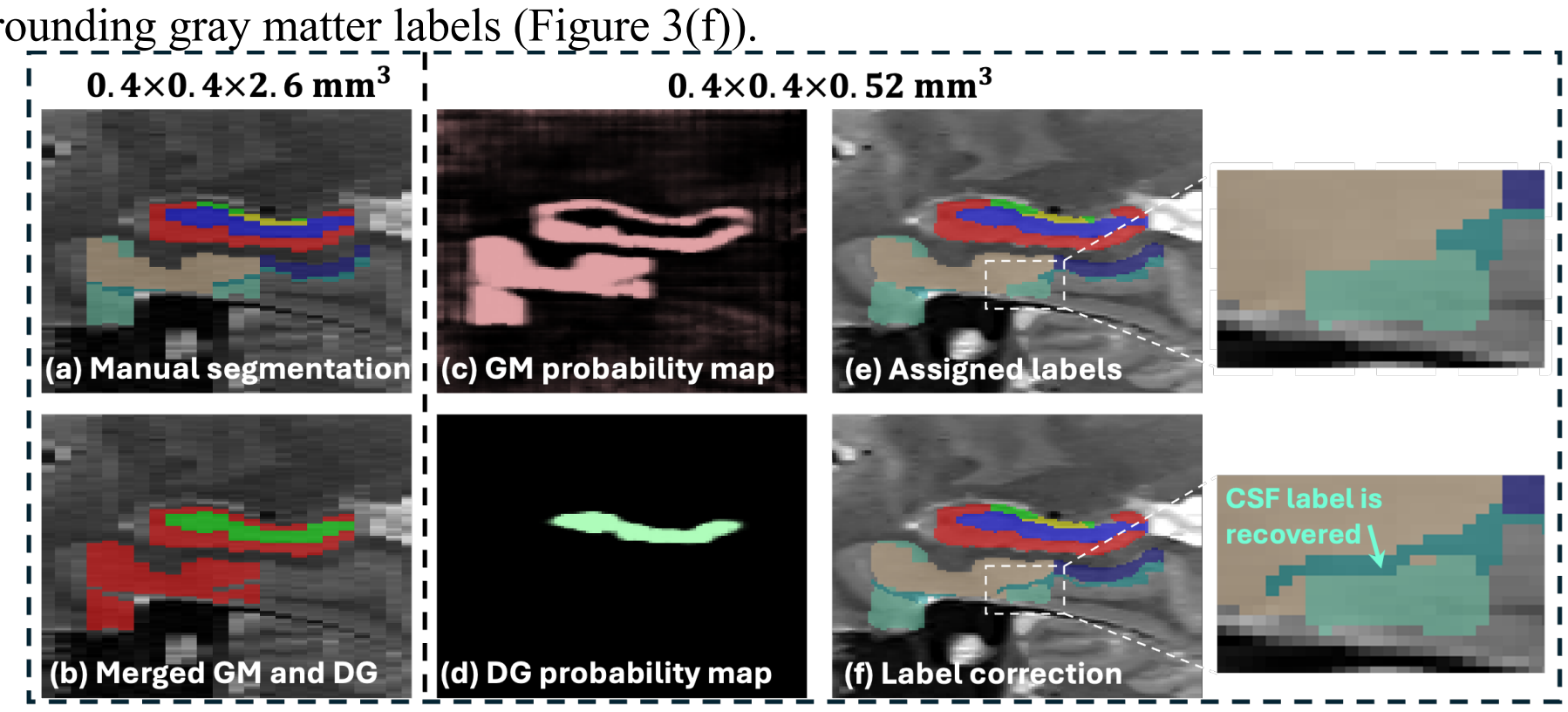

**Figure 3.** Label upsampling pipeline in atlas set. *Abbreviations*: DG, dentate gyrus; GM, gray matter; CSF, cerebrospinal fluid.

After upsampling, a multi-modality segmentation model is trained in the nearly isotropic space. The model uses the nnU-Net framework [16], incorporating multi-modality T1w and T2w images as inputs. Before inputting the T1w image into the network, it is aligned to the T2w image using rigid registration. A modality augmentation strategy is used for the training process. This strategy simulates modality missing scenarios with a 50% probability during training, replacing T1w or T2w images with Gaussian noise, to force the model to extract information from multiple modalities and improve segmentation robustness. The default binary cross-entropy and Dice loss are used for supervision.

### 2.5 Model Evaluation

The trained multi-modality segmentation model was tested on two independent test sets (Figure 1(c)). The T1w and T2w images in the test sets were upsampled to the nearly isotropic space with coronal slice thickness of 0.52 mm using non-local means and linear interpolation, to finish the segmentation.

To evaluate the advantages of the nearly isotropic segmentation model, we reproduced the anisotropic multi-modality method from [11], training and testing it on the same data used in this study to serve as a baseline for comparison

In the cross-sectional test set, the thickness of cortical subregions was extracted from the segmentation results by pruning the Voronoi skeleton of the MTL cortex and integrating the radius field over each subregion [17, 18]. The median thickness of each subregion was used to distinguish between A- CU and A+ MCI using a general linear model with age included as a nuisance covariate. Furthermore, pointwise analysis was also conducted in cross-sectional dataset on a T2w template using a surface-based registration and morphometry pipeline CRASHS [13]. The difference of pointwise thickness between A+ MCI and A- CU was calculated and visualized on a T2w template.

In the longitudinal dataset, the median thickness was compared within each longitudinal pair. Since the participants were A- CU, we expect thickness to remain relatively stable over the 1.4 year mean time window. In the longitudinal paired data, the difference in median thickness for each cortical subregion was calculated. To account for variations in scanning intervals, these differences in median thickness were linearly normalized to a common period of 365 days. Then, the sum of the absolute values of the median thickness differences across the entire dataset was calculated, and the standard deviation of these differences was determined as evaluation metrics. Lower values indicate better longitudinal consistency.

## 3 Results

### 3.1 Visual comparison

Figure 4 shows the results of the multi-modality segmentation model for the same test sample, respectively in anisotropic and nearly isotropic spaces.

The anisotropic model downsampled T1w resolution along the anterior–posterior axis, increasing the slice thickness from 0.8 mm to 1.2 mm in the registration process. This reduced T1w's resolution advantage. Our proposed nearly isotropic model was conducted in upsampled space, maintaining high-resolution properties for both T1w and T2w images.

In Figure 4, the segmentation results in the nearly isotropic space have smoother boundaries than those in the anisotropic space, which is beneficial for the subsequent extraction of imaging biomarkers.

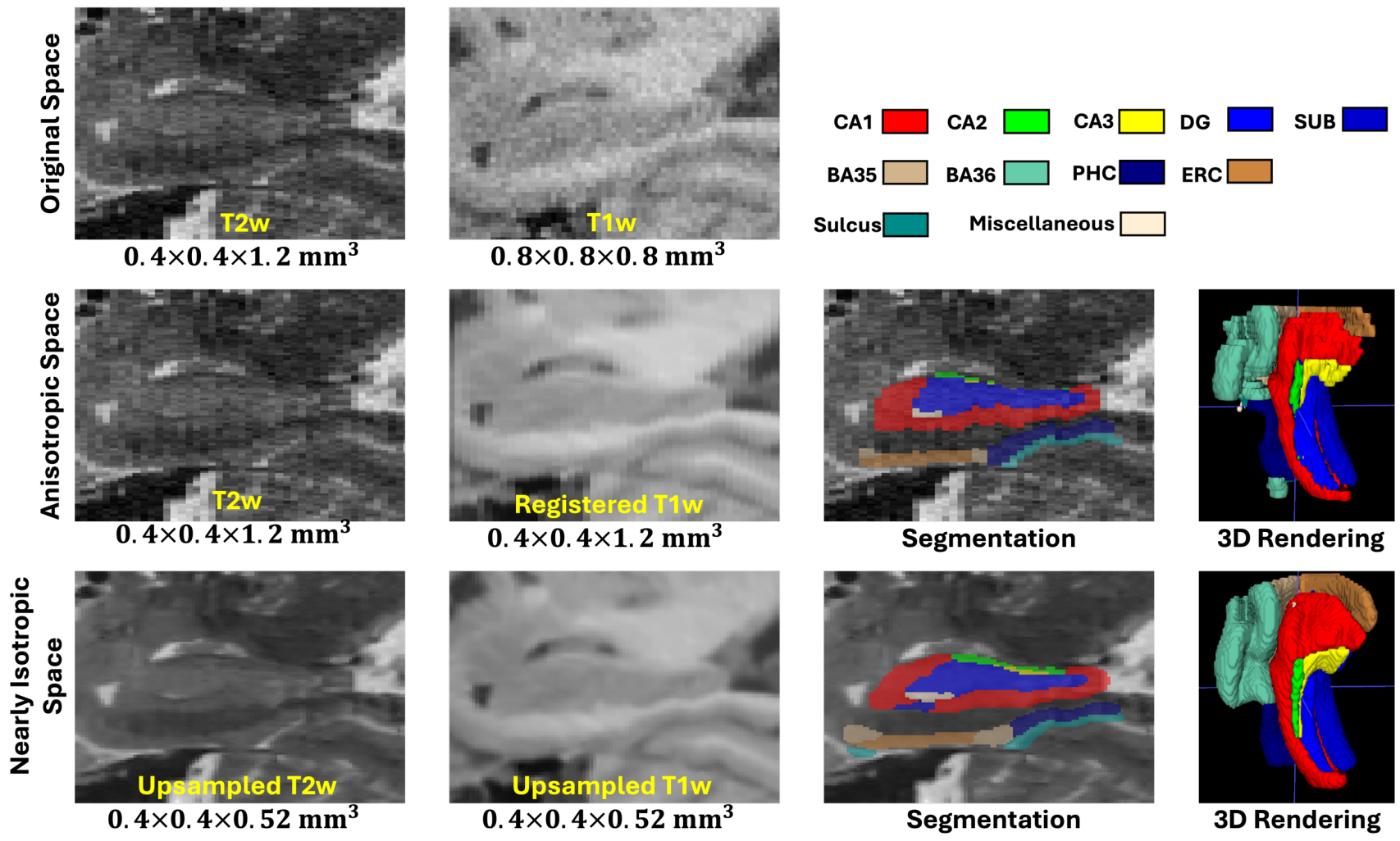

**Figure 4.** Visualization of segmentation. The sagittal view is compared between anisotropic and nearly isotropic spaces

## 3.2 Cross-sectional Evaluation

Table 1 shows the results of fitting the median thickness and the two groups (A- CU and A+ MCI) using a general linear model. The p-value indicates the significance level of the difference in median thickness between A- CU and A+ MCI. The area under the curve (AUC) value reflects the separability of the median thickness between the two groups. A smaller p-value and a higher AUC value represent better discrimination. The findings indicate that imaging biomarkers extracted from nearly isotropic spaces can yield higher AUCs and lower p-values in absolute terms.

**Table 1.** Results of general linear model to distinguish A+ MCI and A- CU groups. The lower p-value and higher AUC are highlighted in each row in blue.

| Space | | Anisotropic Space | | Nearly Isotropic Space | |
|---|---|---|---|---|---|
| | Measures | *p*-value ↓ | AUC ↑ | *p*-value ↓ | AUC ↑ |
| Left | ERC | 7.03E-06 | 0.70 | **4.25E-09** | **0.76** |
| | BA35 | 0.0057 | **0.63** | **0.0022** | **0.63** |
| | BA36 | 4.77E-05 | **0.68** | **4.53E-05** | 0.67 |
| | PHC | 0.74 | 0.49 | **0.034** | **0.60** |
| Right | ERC | 5.51E-05 | 0.68 | **6.59E-06** | **0.70** |
| | BA35 | 0.019 | 0.61 | **0.0032** | **0.64** |
| | BA36 | 0.075 | 0.59 | **0.0072** | **0.62** |
| | PHC | 0.24 | 0.54 | **0.0025** | **0.63** |

↓: the lower the better; ↑: the higher the better.

Figure 5 shows the pointwise comparison between A+ MCI and A- CU groups. The threshold-free cluster enhancement p-value (family-wise error rate corrected) shows that the result in nearly isotropic space has significant differences of pointwise thickness in a larger area between these two groups than that in anisotropic space.

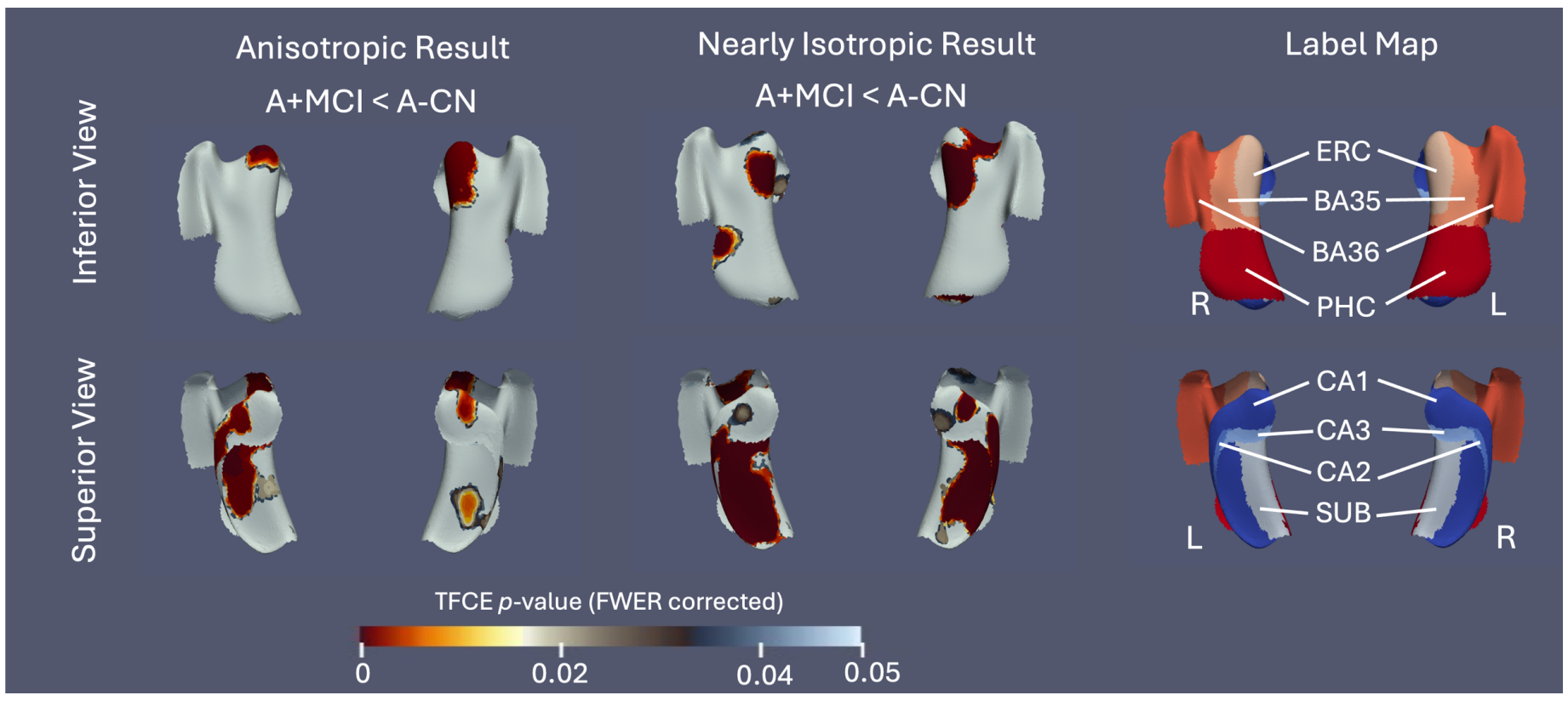

**Figure 5.** The group difference analysis of the point-wise thickness conducted by general linear models in the direction of A+ MCI < A- CU.

### 3.3 Longitudinal Evaluation

Longitudinal thickness consistency across scanning sessions was tested in A- CU individuals, in which MTL subregions were not expected to change significantly in size in the two scanning sessions. Table 2 shows the sum of the absolute values of the median thickness differences across the longitudinal dataset, and the standard deviation of these differences for each cortical subregion.

The biomarkers in nearly isotropic space show better performance in most of the subregions when using absolute sum as measurement. This demonstrates that imaging biomarkers extracted from this space have higher stability.

**Table 2.** Longitudinal thickness consistency analysis in A- CU individuals. The lower standard deviation and sum of the absolute values of thickness longitudinal differences are highlighted in each row in blue.

| Space | | Anisotropic Space | | Nearly Isotropic Space | |
|---|---|---|---|---|---|
| Measures | | Std. ↓ | AbsSum. (mm)↓ | Std. ↓ | AbsSum. (mm)↓ |
| Left | ERC | 0.41 | 5.37 | **0.31** | **4.18** |
| | BA35 | **0.12** | 3.39 | 0.15 | **2.85** |
| | BA36 | 0.37 | 6.13 | **0.31** | **4.49** |
| | PHC | 0.17 | 4.10 | **0.13** | **2.32** |
| Right | ERC | **0.09** | 2.48 | 0.13 | **2.18** |
| | BA35 | 0.25 | 4.45 | **0.21** | **4.12** |
| | BA36 | **0.08** | 2.41 | 0.09 | **2.35** |
| | PHC | **0.14** | **3.40** | 0.30 | 3.89 |

↓: the lower the better; ↑: the higher the better.

## 4 Discussion and conclusion

Using *ex vivo* imaging with 9.4T MRI as a guide, this study established an upsampling model of the MTL subregion atlas set. Based on this, a multi-modality segmentation model for the MTL subregion was developed in a nearly isotropic space. Through downstream tasks, we verified that imaging biomarkers extracted from the isotropic segmentation have better ability to distinguish patients with MCI from individuals who are CU and have stronger longitudinal stability compared to imaging biomarkers in anisotropic spaces.

The nearly isotropic space used in this study allows for a fair analysis of multi-modality images at different scales in two ways. **The first way is within the images of individual participants.** By unifying the voxel spacing in all directions, the calculation of imaging biomarkers achieves a consistent scale across different directions, ensuring that the accuracy of the measurement is independent of direction. **The second way is between different participants.** Due to variations in imaging scanners and acquisition dates, images from different participants or time points may have different voxel resolutions. Directly extracting imaging measures from these images could lead to unfair comparisons between participants. Our proposed nearly isotropic segmentation not only unifies the resolution of multi-modality data from the same participant but also unifies the scale between participants, supporting fair comparison in downstream decision-making in heterogeneous clinical datasets.

Multi-modality data are commonly used by merging them together, such as merging T1w and T2w images in a segmentation model. In this study, we adopt a different approach, using the *ex vivo* modality to provide spatial guidance for the upsampling model rather than merging it directly. In recent years, linking *ex vivo* and *in vivo* data has become a useful approach to discover new *in vivo* imaging biomarkers [19]. Our upsampling network transferred high-resolution structural information from *ex vivo* data into the segmentation model training, making *in vivo* imaging biomarker extraction more effective.

There are two primary limitations in this study. The first limitation is that our proposed space is nearly isotropic rather than real isotropic. In future work, we will explore methods for performing any-scale upsampling to achieve real isotropic segmentation and biomarkers. The second limitation is that the upsampling method can only be applied to manual segmentation in the atlas set, while image upsampling still relies on traditional non-local means, which may cause misalignment during segmentation model training. In future work, we will investigate methods for the simultaneous upsampling of both segmentation and images.

In conclusion, the proposed pipeline can extract imaging biomarkers for neurodegenerative diseases from nearly isotropic multi-modality MR images. They are better at discriminating MCI from cognitively normal individuals and are more robust to temporal changes.

**Acknowledgments.** This work was supported by NIH (grant numbers R01-AG069474, RF1-AG056014, P30-AG072979, R01-AG070592, P01-AG066597, P01-AG084497)

## References

1. Liu, S., Masurkar, A.V., Rusinek, H., Chen, J., Zhang, B., Zhu, W., Fernandez-Granda, C., Razavian, N.: Generalizable deep learning model for early Alzheimer's disease detection from structural MRIs. Sci. Rep. 12, 17106 (2022). https://doi.org/10.1038/s41598-022-20674-x.
2. Thompson, P.M., Hayashi, K.M., Dutton, R.A., Chiang, M., Leow, A.D., Sowell, E.R., De Zubicaray, G., Becker, J.T., Lopez, O.L., Aizenstein, H.J., Toga, A.W.: Tracking Alzheimer's Disease. Ann. N. Y. Acad. Sci. 1097, 183–214 (2007). https://doi.org/10.1196/annals.1379.017.
3. De Flores, R., Wisse, L.E.M., Das, S.R., Xie, L., McMillan, C.T., Trojanowski, J.Q., Robinson, J.L., Grossman, M., Lee, E., Irwin, D.J., Yushkevich, P.A., Wolk, D.A.: Contribution of mixed pathology to medial temporal lobe atrophy in Alzheimer's disease. Alzheimers Dement. 16, 843–852 (2020). https://doi.org/10.1002/alz.12079.
4. Nelson, P.T., Dickson, D.W., Trojanowski, J.Q., Jack, C.R., Boyle, P.A., Arfanakis, K., Rademakers, R., Alafuzoff, I., Attems, J., Brayne, C., Coyle-Gilchrist, I.T.S., Chui, H.C., Fardo, D.W., Flanagan, M.E., Halliday, G., Hokkanen, S.R.K., Hunter, S., Jicha, G.A., Katsumata, Y., Kawas, C.H., Keene, C.D., Kovacs, G.G., Kukull, W.A., Levey, A.I., Makkinejad, N., Montine, T.J., Murayama, S., Murray, M.E., Nag, S., Rissman, R.A., Seeley, W.W., Sperling, R.A., White Iii, C.L., Yu, L., Schneider, J.A.: Limbic-predominant age-related TDP-43 encephalopathy (LATE): consensus working group report. Brain. 142, 1503–1527 (2019). https://doi.org/10.1093/brain/awz099.
5. Yushkevich, P.A., Pluta, J.B., Wang, H., Xie, L., Ding, S.-L., Gertje, E.C., Mancuso, L., Kliot, D., Das, S.R., Wolk, D.A.: Automated volumetry and regional thickness analysis of hippocampal subfields and medial temporal cortical structures in mild cognitive impairment: Automatic Morphometry of MTL Subfields in MCI. Hum. Brain Mapp. 36, 258–287 (2015). https://doi.org/10.1002/hbm.22627.
6. Fischl, B.: FreeSurfer. NeuroImage. 62, 774–781 (2012). https://doi.org/10.1016/j.neuroimage.2012.01.021.
7. DeKraker, J., Haast, R.A., Yousif, M.D., Karat, B., Lau, J.C., Köhler, S., Khan, A.R.: Automated hippocampal unfolding for morphometry and subfield segmentation with HippUnfold. eLife. 11, e77945 (2022). https://doi.org/10.7554/eLife.77945.
8. De Flores, R., La Joie, R., Chételat, G.: Structural imaging of hippocampal subfields in healthy aging and Alzheimer's disease. Neuroscience. 309, 29–50 (2015). https://doi.org/10.1016/j.neuroscience.2015.08.033.
9. Ver Hoef, L., Deshpande, H., Cure, J., Selladurai, G., Beattie, J., Kennedy, R.E., Knowlton, R.C., Szaflarski, J.P.: Clear and Consistent Imaging of Hippocampal Internal Architecture With High Resolution Multiple Image Co-registration and Averaging (HR-MICRA). Front. Neurosci. 15, 546312 (2021). https://doi.org/10.3389/fnins.2021.546312.
10. Xie, L., Wisse, L.E.M., Pluta, J., De Flores, R., Piskin, V., Manjón, J.V., Wang, H., Das, S.R., Ding, S., Wolk, D.A., Yushkevich, P.A., for the Alzheimer's Disease Neuroimaging Initiative: Automated segmentation of medial temporal lobe subregions on *in vivo* T1-weighted MRI in early stages of Alzheimer's disease. Hum. Brain Mapp. 40, 3431–3451 (2019). https://doi.org/10.1002/hbm.24607.

11. Li, Y., Xie, L., Khandelwal, P., Wisse, L.E.M., Brown, C.A., Prabhakaran, K., Tisdall, M.D., Mechanic-Hamilton, D., Detre, J.A., Das, S.R., Wolk, D.A., Yushkevich, P.A.: Automatic segmentation of medial temporal lobe subregions in multi-scanner, multi-modality MRI of variable quality. Bioengineering (2024). https://doi.org/10.1101/2024.05.21.595190.
12. Ravikumar, S., Denning, A.E., Lim, S., Chung, E., Sadeghpour, N., Ittyerah, R., Wisse, L.E.M., Das, S.R., Xie, L., Robinson, J.L., Schuck, T., Lee, E.B., Detre, J.A., Tisdall, M.D., Prabhakaran, K., Mizsei, G., De Onzono Martin, M.M.I., Arroyo Jiménez, M.D.M., Mũnoz, M., Marcos Rabal, M.D.P., Cebada Sánchez, S., Delgado González, J.C., De La Rosa Prieto, C., Irwin, D.J., Wolk, D.A., Insausti, R., Yushkevich, P.A.: Postmortem imaging reveals patterns of medial temporal lobe vulnerability to tau pathology in Alzheimer's disease. Nat. Commun. 15, 4803 (2024). https://doi.org/10.1038/s41467-024-49205-0.
13. Yushkevich, P.A., Ittyerah, R., Li, Y., Denning, A.E., Sadeghpour, N., Lim, S., McGrew, E., Xie, L., DeFlores, R., Brown, C.A., Wisse, L.E.M., Wolk, D.A., Das, S.R., for the Alzheimer's Disease Neuroimaging Initiative: Morphometry of medial temporal lobe subregions using high-resolution T2-weighted MRI in ADNI3: Why, how, and what's next? Alzheimers Dement. 20, 8113–8128 (2024). https://doi.org/10.1002/alz.14161.
14. Ronneberger, O., Fischer, P., Brox, T.: U-Net: Convolutional Networks for Biomedical Image Segmentation. In: Navab, N., Hornegger, J., Wells, W.M., and Frangi, A.F. (eds.) Medical Image Computing and Computer-Assisted Intervention – MICCAI 2015. pp. 234–241. Springer International Publishing, Cham (2015). https://doi.org/10.1007/978-3-319-24574-4_28.
15. Manjón, J.V., Coupé, P., Buades, A., Fonov, V., Louis Collins, D., Robles, M.: Non-local MRI upsampling. Med. Image Anal. 14, 784–792 (2010). https://doi.org/10.1016/j.media.2010.05.010.
16. Isensee, F., Jaeger, P.F., Kohl, S.A.A., Petersen, J., Maier-Hein, K.H.: nnU-Net: a self-configuring method for deep learning-based biomedical image segmentation. Nat. Methods. 18, 203–211 (2021). https://doi.org/10.1038/s41592-020-01008-z.
17. Ogniewicz, R.L., Kübler, O.: Hierarchic Voronoi Skeletons. Pattern Recognit. 28, 343–359 (1995). https://doi.org/doi.org/10.1016/0031-3203(94)00105-U.
18. Pouch, A.M., Tian, S., Takebe, M., Yuan, J., Gorman, R., Cheung, A.T., Wang, H., Jackson, B.M., Gorman, J.H., Gorman, R.C., Yushkevich, P.A.: Medially constrained deformable modeling for segmentation of branching medial structures: Application to aortic valve segmentation and morphometry. Med. Image Anal. 26, 217–231 (2015). https://doi.org/10.1016/j.media.2015.09.003.
19. Tian, L., Young, S.I., Ramirez, J.W., Zemlyanker, D., Binder, L.J.D., Herisse, R., Connors, T.R., Oakley, D.H., Hyman, B.T., Puonti, O., Rosen, M.S., Iglesias, J.E.: Reference-Free 3D Reconstruction of Brain Dissection Photographs with Machine Learning, http://arxiv.org/abs/2503.09963, (2025).